\documentclass[twocolumn,prb,showpacs,preprintnumbers,amsmath,amssymb,superscriptaddress]{revtex4}

\usepackage{graphicx}
\usepackage{dcolumn}
\usepackage{bm}

\begin{document}


\title{Two-orbital Kondo effect in quantum dot coupled to ferromagnetic leads}

\author{Hitoshi Yoshizumi}
\affiliation{Department of Applied Physics, Osaka University, Suita, Osaka 565-0871, Japan.}
\author{Kensuke Inaba}
\affiliation{NTT Basic Research Laboratories, NTT Corporation, Atsugi 243-0198, Japan}%
\affiliation{JST, CREST, Chiyoda-ku, Tokyo 102-0075, Japan}%
\author{Tomoko Kita}
\thanks{Present address: Department of Physics, Kyoto University, Kyoto 606-8502, Japan.}
\affiliation{Department of Applied Physics, Osaka University, Suita, Osaka 565-0871, Japan.}%
\author{Sei-ichiro Suga}
\affiliation{Department of Applied Physics, Osaka University, Suita, Osaka 565-0871, Japan.}
\affiliation{Department of Materials Science and Chemistry, University of Hyogo, Himeji, Hyogo 671-2280, Japan}
\date{\today}

\begin{abstract}
We study the Kondo effect of a two-orbital vertical quantum dot (QD) coupled to two ferromagnetic leads by employing an equation of motion method. When the ferromagnetic leads are coupled with parallel spin polarization, we find three peaks in the single-particle excitation spectra. The middle one is the Kondo resonance caused by the orbital degrees of freedom. 
In magnetic fields, the Kondo effect vanishes. However, at a certain magnetic field new two-fold degenerate states arise and the Kondo effect emerges there. 
In contrast, when the ferromagnetic leads are coupled with antiparallel spin polarization, the Kondo effect caused by the spin (orbital) degrees of freedom survives (is suppressed) in magnetic fields. 
We investigate the field dependence of the conductance in the parallel and antiparallel spin polarizations of the leads and find that the conductance changes noticeably in magnetic fields. 
\end{abstract}

\pacs{72.15.Qm, 73.63.Kv, 72.25.-b, 85.75.-d}
\maketitle

\section{Introduction}
The Kondo effect, which was originally studied in dilute magnetic alloys, is a typical phenomenon caused by local electron correlation. Recent progress on nano-processing techniques has enabled us to observe the Kondo effect in quantum dot (QD) systems \cite{qd1,qd2}. Since the QD system has many tunable parameters, various aspects of the Kondo effect can be controlled. 
For instance, the multiorbital Kondo effect realized in vertical QDs \cite{vqd1,vqd2} and carbon-nanotube QDs \cite{cntqd,makarovski07} has attracted much interest. 
In particular, in a vertical QD, the confinement potential yields multiply degenerate single-particle QD energy levels called the Fock-Darwin states \cite{fock28,darwin30,tarucha96-97}, which are regarded as effective orbital degrees of freedom. 
The degeneracy of the QD states in the confinement potential can be controlled with external magnetic fields. 
The introduction of the orbital splitting of the QD energy levels realizes a crossover from the SU(4) to the SU(2) Kondo effect, which results in characteristic transport properties \cite{sasaki2005,eto2005,lim06,choi05,sakano06}. In the SU(4) Kondo effect, a higher Kondo temperature than that in the SU(2) Kondo effect was observed in experiments \cite{vqd2,cntqd,makarovski07}.

Spin-dependent current in metallic leads is another interesting topic in relation to QD systems. Recently, there have been a number of experimental studies of QD systems coupled to ferromagnetic leads in the context of spintronics \cite{ferroC60,ferroInAs,ferroInAs1,ferroCNT}. 
The splitting and restoration of the Kondo peak has been observed in experiments on a $\rm{C_{60}}$ QD system \cite{ferroC60} and a semiconductor QD system, which are coupled to ferromagnetic leads \cite{ferroInAs,ferroInAs1}. 
There have been extensive theoretical studies on a single-orbital QD coupled to two ferromagnetic leads \cite{lopez03,ferroEOM1,zhang09,ferroNRG,ferrotheta1,ferrotheta2,choi07,utsumi05,martinek05,polish06,polish07,sindel07}. 
When the spin polarization of the two ferromagnetic leads is parallel, the Kondo effect caused by the spin degrees of freedom is suppressed even in the absence of magnetic fields. On the other hand, the Kondo effect remains when the spin polarization of the two ferromagnetic leads is antiparallel \cite{ferrotheta1,ferrotheta2,polish06}. 
These results motivated us to investigate a multiorbital QD coupled to ferromagnetic leads. 
One can expect that the interplay between the spin polarization and the multiorbital effects causes characteristic features of the Kondo effect and the related transport properties. 
For example, in a two-orbital QD coupled to ferromagnetic leads, the appearance of the underscreening Kondo effect was discussed \cite{borda}. In carbon nanotube QD, the influence of ferromagnetic leads on the shot noise was analyzed \cite{polish10}.

In this paper, we investigate the Kondo effect of a two-orbital vertical QD coupled to two ferromagnetic leads. 
By employing an equation of motion (EOM) method \cite{ferroEOM1,polish06,polish07,zhang09,EOM1,EOM2}, we calculate the single-particle excitation spectra (SPES) of the localized electron in the QD and the conductance. The SPES for several spin polarization values are discussed in connection with the splitting of the QD energy levels obtained by the poor man's scaling approach \cite{ferrotheta1,ferrotheta2,haldane78}. 
We also investigate the effects of the orbital splitting caused by magnetic fields on the SPES. 
The conductance is shown as a function of the orbital splitting.

This paper is organized as follows. 
In Sec. \ref{sec:m&m}, we outline the model and the methods. We consider a QD with two orbitals coupled to two ferromagnetic leads. The system is modeled as a two-orbital Anderson impurity Hamiltonian where the conduction electrons have a spin dependent density of states (DOS). We outline an EOM method and a poor man's scaling approach. 
In Sec. \ref{sec:res}, we show the results. In the SPES for the parallel spin polarization of the leads, we find that three peaks appear and the middle one is located close to the Fermi energy. It is considered that this is a Kondo resonance caused by the orbital degrees of freedom. 
When a magnetic field is applied, the energy levels of the QD split and the Kondo effect disappears. However, in a certain magnetic field two energy levels cross and the Kondo effect emerges again. 
We discuss the conductance as a function of the energy-level splitting of the QD. A summary is provided in Sec. \ref{sec:sum}.

\section{Model and methods} 
\label{sec:m&m}
\subsection{Two-orbital quantum dot coupled to ferromagnetic leads}
\begin{figure}[!t]
\begin{center}
\includegraphics[width=75mm]{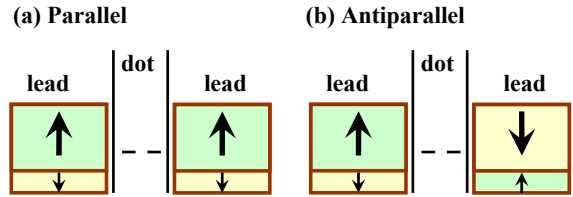}
\end{center}
\caption{\baselineskip=0.77\baselineskip
(Color on-line) Schematic diagram of the magnetic configurations in our QD system. The relative orientations of the magnetic moments in the two ferromagnetic leads are (a) parallel and (b) antiparallel.}
\label{fig1}
\end{figure}

We consider a single QD with two-orbital degrees of freedom coupled to two ferromagnetic leads. The system is described by the following Hamiltonian, 
\begin{eqnarray}
H &=& \sum_{i = L,R} \sum_{k,l,\sigma} \varepsilon_{i k \sigma} c_{i k l \sigma}^\dagger c_{i k l \sigma}
 + \sum_{l,\sigma} \varepsilon_{l} d_{l \sigma}^\dagger d_{l \sigma}\nonumber\\
 &+& \sum_{i = L,R} \sum_{k,l,\sigma} (V_{i k l \sigma} c_{i k l \sigma}^\dagger d_{l \sigma} + h.c. )\nonumber\\
 &+& U \sum_{l} n_{l \uparrow} n_{l \downarrow}
 + U^\prime \sum_{\sigma \sigma^\prime} n_{l \sigma} n_{\overline{l} \sigma^\prime},
\end{eqnarray}
where $c_{i k l \sigma}^{(\dagger)}$ annihilates (creates) a conduction electron with wave number $k$, spin $\sigma$, and orbital $l$ in the lead $i$ (=$L$, $R$), $\varepsilon_{l}$ is the QD energy level for orbital $l$, $d_{l \sigma}^{(\dagger)}$ annihilates (creates) a localized electron with spin $\sigma$ and orbital $l$ in the QD, and $n_{l \sigma} = d_{l \sigma}^{\dagger} d_{l \sigma}$. Intraorbital and interorbital Coulomb interactions are expressed by $U$ and $U^\prime$, respectively, and 
$V_{i k l \sigma}$ is the tunneling amplitude between the leads and the QD. 
We set $U$ and $U' \gg |\varepsilon_l|$ as realized in many QD systems, so that the double occupancy of the QD electron is forbidden and the electron number is at most unity. Therefore, we can neglect the effects of the exchange interaction. 
We assume that the orbital states in the QD hybridize with the corresponding conduction channels in the leads. 
Although the assumption of multiple conduction channels is nontrivial,
it is known that this is relevant for certain systems, e.g. vertical QD systems and carbon nanotube QD systems \cite{vqd1,cntqd,vqd2}. 
The spin polarization of the ferromagnetic leads is defined by $p = ( \rho_{c i \uparrow} - \rho_{c i \downarrow} ) / ( \rho_{c i \uparrow} + \rho_{c i \downarrow} )$, ($0 \le p \le 1$), where $\rho_{c i \sigma}$ is the spin dependent DOS of the conduction electron in the lead $i$ (=$L$, $R$), which is assumed to be independent of orbital $l$. 
We use the energy-independent constant $\rho_{c i \sigma}$, which is reasonable in the wide-band limit. 
For a parallel magnetic (P) configuration, the relative orientation of the magnetic moments in the two ferromagnetic leads is parallel as shown in Fig. \ref{fig1}(a). Both DOSs in the leads $L$ and $R$ are defined by $\rho_{c L \sigma} = \rho_{c R \sigma} = \left[1 + (\delta_{\sigma, \uparrow} - \delta_{\sigma, \downarrow}) p \right] \rho_0$ with $\rho_{0}$ being the DOS of the leads at $p=0$. 
For an antiparallel magnetic (AP) configuration, as shown in Fig. \ref{fig1}(b), the DOSs in the leads $L$ and $R$ are defined by $\rho_{c L \sigma} = \left[1 + (\delta_{\sigma, \uparrow} - \delta_{\sigma, \downarrow}) p \right] \rho_0$ and $\rho_{c R \sigma} = \left[1 - (\delta_{\sigma, \uparrow} - \delta_{\sigma, \downarrow}) p \right] \rho_0$, respectively.

In a vertical QD system, the single-particle energy level of the QD in a magnetic field can be described as the Fock-Darwin state. 
The energy levels of two orbitals are split into two levels, $\varepsilon_{1}$ and $\varepsilon_{2}$, in magnetic fields. We model this splitting as 
\begin{eqnarray}
\varepsilon_{l} = \varepsilon_{0} + (\delta_{l,1} - \delta_{l,2}) \Delta_{orb}, \hspace{0.5cm} (l = 1,2),
\end{eqnarray}
where $\varepsilon_{0}$ is the center of the QD energy level and $\Delta_{orb}$ is the splitting width. We call $\Delta_{orb}$ the orbital splitting in the following. 
Because of the small $g$ factor in semiconductors, the Zeeman splitting of the semiconductor QD system is much smaller than $\Delta_{orb}$ in a magnetic field, so that we can ignore the effects of the Zeeman splitting \cite{sasaki2005}. 
We thus obtain the field dependence via the $\Delta_{orb}$ dependence using the Fock-Darwin energy diagram \cite{tarucha96-97}.

\subsection{Equation of motion method}
We calculate the retarded Green's function $G_{l \sigma}^r (t) = - i \theta (t) \langle \{ d_{l \sigma} (t), d_{l \sigma}^\dagger (0) \} \rangle$ for a localized electron of the QD. For this purpose, we employ an EOM method. 
We perform the straightforward EOM calculations until the second iterative procedure. In the third iterative EOM procedure, we use the approximation which successfully yields the closed equations of the higher-order Green's functions in a single-orbital QD system \cite{EOM1,EOM2}. In addition to this approximation, we include the higher-order terms expressing the scattering processes of the electrons in the same orbital with different spins, and in the different orbitals with the same and different spins. We neglect the higher-order terms expressing other higher-order scattering processes. 
Details of the calculations are described in Appendix.

In the condition of the strong interactions $U$ and $U' \gg |\varepsilon_l|$, we obtain the equations for the Green's function as
\begin{eqnarray}
G^r_{l \sigma} (\omega) = \frac{ 1
 - \langle n_{l \overline{\sigma}} \rangle
 - \langle n_{\overline{l} \sigma} \rangle
 - \langle n_{\overline{l} \overline{\sigma}} \rangle }
{\omega - \varepsilon_{l}
 - \Sigma^0_{l \sigma} (\omega)
 - \Sigma^1_{l \sigma} (\omega)
 - \Sigma^2_{l \sigma} (\omega)
 - \Sigma^3_{l \sigma} (\omega)}, 
\nonumber
\label{eq:gf}
\end{eqnarray}
\vspace{-0.6cm}
\begin{eqnarray}
\end{eqnarray}
where
\begin{eqnarray}
\Sigma^0_{l \sigma} (\omega) = \sum_{i = L,R} \sum_{k} | V_{i k l \sigma} |^2 \frac{ 1 }{\omega - \varepsilon_{i k \sigma}},
\label{eq:se0}
\end{eqnarray}
\begin{eqnarray}
\Sigma^1_{l \sigma} (\omega) = \sum_{i = L,R} \sum_{k} | V_{i k l \overline{\sigma}} |^2 f (\varepsilon_{i k \overline{\sigma}}) \frac{ 1 }{\omega - \varepsilon_{i k \overline{\sigma}} + \Delta \tilde{\varepsilon}^1_{l \sigma}},
\label{eq:se1}
\end{eqnarray}
\begin{eqnarray}
\Sigma^2_{l \sigma} (\omega) = \sum_{i = L,R} \sum_{k} | V_{i k \overline{l} \sigma} |^2 f (\varepsilon_{i k \sigma}) \frac{ 1 }{\omega - \varepsilon_{i k \sigma} + \Delta \tilde{\varepsilon}^2_{l \sigma}},
\label{eq:se2}
\end{eqnarray}
\begin{eqnarray}
\Sigma^3_{l \sigma} (\omega) = \sum_{i = L,R} \sum_{k} | V_{i k \overline{l} \overline{\sigma}} |^2 f (\varepsilon_{i k \overline{\sigma}}) \frac{ 1 }{\omega - \varepsilon_{i k \overline{\sigma}} + \Delta \tilde{\varepsilon}^3_{l \sigma}},
\label{eq:se3}
\end{eqnarray}
with $\overline{\sigma}$ being the opposite spin state to $\sigma$ and $\overline{l}$ being the different orbital state from $l$. 
The Fermi energy is set at $\omega=0$. 
The self-energy $\Sigma^1_{l \sigma} (\omega)$ expresses the scattering process of the electrons in the same orbital with different spins. 
The self-energies $\Sigma^2_{l \sigma} (\omega)$ and $\Sigma^3_{l \sigma} (\omega)$ express the scattering processes of the electrons in the different orbitals with the same and different spins, respectively. 
Accordingly, spin fluctuations are introduced via $\Sigma^1_{l \sigma} (\omega)$ and $\Sigma^3_{l \sigma} (\omega)$, while orbital fluctuations are introduced via $\Sigma^2_{l \sigma} (\omega)$ and $\Sigma^3_{l \sigma} (\omega)$. 
These scattering processes play essential roles in the appearance of the Kondo effect caused by spin and orbital fluctuations. 
We thus argue that the present EOM approximation captures the essentials of the Kondo effect in the two-orbital QD.

The electron number $\langle n_{l \sigma} \rangle$ is determined using the equation, 
\begin{eqnarray}
\langle n_{l \sigma} \rangle = - \frac{1}{\pi} \int_{}^{} \mathrm{Im} [G^r_{l \sigma} (\omega)] f(\omega) \,d \omega . 
\label{eq:n}
\end{eqnarray}
In this way, we complete the self-consistent loop which consists of Eqs. (\ref{eq:gf})-(\ref{eq:n}).

To improve the numerical results quantitatively, we have further introduced the second self-consistent loop. 
By analogy with the self-consistent procedure used for a single-orbital QD coupled to ferromagnetic leads \cite{ferroEOM1}, we have replaced the bare QD energy splittings in $\Sigma^1_{l \sigma} (\omega)$, $\Sigma^2_{l \sigma} (\omega)$ and $\Sigma^3_{l \sigma} (\omega)$ by the renormalized ones $\Delta \tilde{\varepsilon}^1_{l \sigma}$, $\Delta \tilde{\varepsilon}^2_{l \sigma}$ and $\Delta \tilde{\varepsilon}^3_{l \sigma}$, respectively, as 
\begin{eqnarray}
\Delta \tilde{\varepsilon}^1_{l \sigma} = \tilde{\varepsilon}_{l \overline{\sigma}} - \tilde{\varepsilon}_{l \sigma},
\label{eq:del_e1}
\end{eqnarray}
\begin{eqnarray}
\Delta \tilde{\varepsilon}^2_{l \sigma} = \tilde{\varepsilon}_{\overline{l} \sigma} - \tilde{\varepsilon}_{l \sigma},
\label{eq:del_e2}
\end{eqnarray}
\begin{eqnarray}
\Delta \tilde{\varepsilon}^3_{l \sigma} = \tilde{\varepsilon}_{\overline{l} \overline{\sigma}} - \tilde{\varepsilon}_{l \sigma},
\label{eq:del_e3}
\end{eqnarray}
\begin{eqnarray*}
\tilde{\varepsilon}_{l \sigma} = \varepsilon_{l}
 + \mathrm{Re} [ \Sigma^0_{l \sigma} (\tilde{\varepsilon}_{l \sigma})
 + \Sigma^1_{l \sigma} (\tilde{\varepsilon}_{l \sigma})
 + \Sigma^2_{l \sigma} (\tilde{\varepsilon}_{l \sigma})
 + \Sigma^3_{l \sigma} (\tilde{\varepsilon}_{l \sigma})
], 
\label{eq:renor}
\end{eqnarray*}
\vspace{-0.8cm}
\begin{eqnarray}
\end{eqnarray}
where $\Delta \tilde{\varepsilon}^1_{l \sigma}$ denotes the difference between the QD levels of the same orbital with different spins, and 
$\Delta \tilde{\varepsilon}^2_{l \sigma}$ and $\Delta \tilde{\varepsilon}^3_{l \sigma}$ denote the difference between the QD levels of a different orbital with the same and different spins, respectively. 
At $\omega=\tilde{\varepsilon}_{l \sigma}$ determined self-consistently in Eq. (\ref{eq:renor}), the real part of the denominator of the Green's function in Eq. (\ref{eq:gf}) becomes zero. This fact suggests that the additional self-consistent procedure provides us with the important information about the renormalized energy level of the electrons in the QD. 
For the single-orbital case in the P configuration, the splitting of the Kondo resonance was obtained using this EOM method \cite{ferroEOM1}. This feature was confirmed by using the numerical renormalization-group calculation \cite{ferroNRG}.
The splitting of the Kondo resonance caused by the ferromagnetic leads is often characterized by the so-called local exchange field \cite{martinek05}.

Substituting the self-consistently obtained Green's function into the relation 
\begin{eqnarray}
\rho_{l \sigma}(\omega) = - \frac{1}{\pi} \mathrm{Im} \left[ G^r_{l \sigma} (\omega) \right], 
\end{eqnarray}
we obtain the SPES for the localized electron of the QD. 
Note that the spectral weights obtained by the EOM method are rather underestimated and are quantitatively not accurate, although the spectral profiles are qualitatively correct.  
The bias-linear conductance is expressed as \cite{EOM2,meir92,meir94} 
$
G = -(2 \pi e^2/h) \int d \omega f^\prime(\omega)
\sum _{l \sigma} \rho _{l \sigma}(\omega) 
\Gamma^L_{l \sigma} \Gamma^R_{l \sigma}/ 
\left( \Gamma^L_{l \sigma} + \Gamma^R_{l \sigma} \right)
$, 
where $f^\prime(\omega) = \partial f(\omega)/\partial \omega$ with $f(\omega)$ being the Fermi distribution function and $\Gamma_{l \sigma}^{L(R)}$ is the hybridization between the left (right) lead and the QD. 
Note that this expression for the conductance is valid, if the Green's function $G^r_{l \sigma} (\omega)$ is diagonalized with respect to $l$ and $\sigma$. 
When we consider the QD system with off-diagonal contributions as well as plural electrons, we have to also improve the present EOM formulation. Such a situation is beyond the current interest.

In the calculation, we set a flat band and assume that $V_{i k l \sigma}$ is independent of $i$, $k$, $l$, and $\sigma$ as $V_{i k l \sigma} = V$. By these reasonable simplifications, we obtain 
$\Gamma_{l \sigma}^L=\left[1 + (\delta_{\sigma \uparrow} - \delta_{\sigma \downarrow})p \right] \Gamma_{0}/2$ and 
$\Gamma_{l \sigma}^R=\left[1 \pm (\delta_{\sigma \uparrow} - \delta_{\sigma \downarrow})p \right] \Gamma_{0}/2$, where $\pm$ denotes the P($+$) and AP($-$) configurations, and $\Gamma_{0}=8\pi \rho_0 V^2$ is the total hybridization of the left and right leads at $p=0$. 
The conductance takes the form 
\begin{eqnarray}
G &=& - 2 \pi \frac{e^2 }{h }
\int d \omega f^\prime(\omega) \frac{\Gamma_0}{4} \sum _{l \sigma} 
              \rho _{l \sigma}(\omega) 
\nonumber \\
&\times& \begin{cases}
         \left[1+\left(\delta_{\sigma \uparrow}-
                       \delta_{\sigma \downarrow} \right)p \right], & 
                       {\rm P \: configuration}, \\
         \left[1-p^2 \right], & 
                       {\rm AP \: configuration}. 
         \end{cases} 
\label{eq:G}
\end{eqnarray}

\subsection{Poor man's scaling approach}
We evaluate the splitting of the renormalized QD energy levels for the P configuration using a poor man's scaling approach \cite{ferrotheta2,ferrotheta1} which is known as a powerful tool to study the renormalization effect in the Anderson model \cite{haldane78}.
We reduce the energy scale of the half width of the flat band $D$ to $D_{1}$, where charge fluctuations are quenched. By integrating out the excitations in the energy range between $D_{1}$ and $D$, we obtain the renormalized QD energy levels $\tilde{\varepsilon}^{scal}_{l \uparrow}$ and $\tilde{\varepsilon}^{scal}_{l \downarrow}$. 
Because of this renormalization, the degenerate two energy levels yield the spin splitting, which is given by
\begin{eqnarray}
\Delta \tilde{\varepsilon}_{scal} \equiv \tilde{\varepsilon}^{scal}_{l \downarrow} - \tilde{\varepsilon}^{scal}_{l \uparrow} = \frac{1}{\pi} p \Gamma_{0} \mathrm{ln} \frac{D}{D_{1}},
\label{eq:de_scal}
\end{eqnarray}
where $D_{1}$ is determined from the condition that the lower band edge $-D_1$ coincides with the lower renormalized QD level $\tilde{\varepsilon}^{scal}_{l \uparrow}$ according to the procedure in Refs. \cite{ferrotheta2,ferrotheta1}: 
\begin{eqnarray}
-D_{1} = \varepsilon_{0} + \frac{1}{2 \pi} (1-p) \Gamma_{0} \mathrm{ln} \frac{D}{D_{1}}.
\label{eq:D1_scal}
\end{eqnarray}
We obtain $D_1$ for given $p$ and $D$ from Eq. (\ref{eq:D1_scal}). Substituting the obtained $D_1$ into Eq. (\ref{eq:de_scal}), we determine the $p$ dependence of $\Delta \tilde{\varepsilon}_{scal}$.

\section{Results and discussions} \label{sec:res}
On the basis of the formulations obtained in Sec. \ref{sec:m&m}, we carry out the numerical calculations, keeping a vertical QD and Ni ferromagnetic leads in mind. We use $\Gamma_{0}$ in units of energy. 
We adopt $D/\Gamma_{0}=100$, $\varepsilon_{0}/\Gamma_{0} = -2$, and temperature $T/\Gamma_{0}=0.005$. Since the spin polarization of the Ni ferromagnetic lead is $p \sim 0.3$ \cite{ferroC60}, we direct our attention to the $p=0.3$ system and the $p$ dependence up to around $p=0.3$.

\subsection{Spin polarization dependence of Kondo effect}
\renewcommand{\figurename}{Fig.}
\begin{figure}[!t]
\begin{center}
\includegraphics[width=70mm]{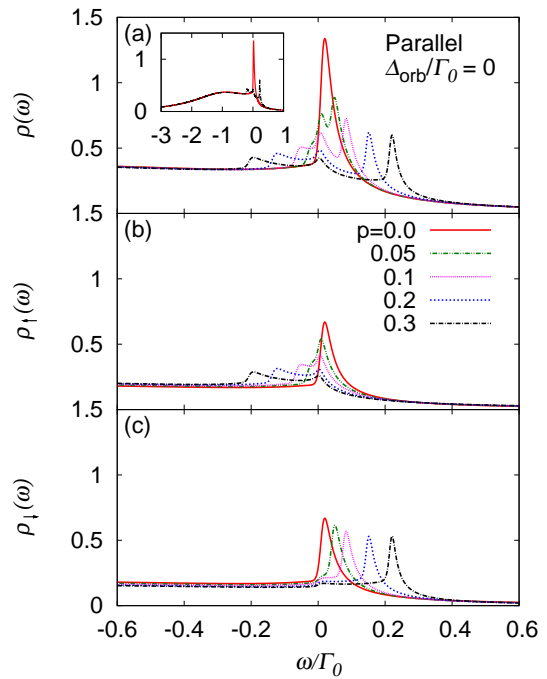}
\end{center}
\caption{\baselineskip=0.77\baselineskip
(Color on-line) Single-particle excitation spectra (a) $\rho (\omega) = \rho_{\uparrow} (\omega) + \rho_{\downarrow} (\omega)$, (b) $\rho_{\uparrow} (\omega)$, and (c) $\rho_{\downarrow} (\omega)$ for several $p$ values in the P configuration. 
The parameters are $\Delta_{orb}/\Gamma_{0} = 0$, $T/\Gamma_{0} = 0.005$, and $\varepsilon_{0}/\Gamma_{0} = -2$. 
Inset: Expanded scale of $\rho (\omega)$. 
}
\label{fig2}
\end{figure}
\begin{figure}[!t]
\begin{center}
\includegraphics[width=\linewidth]{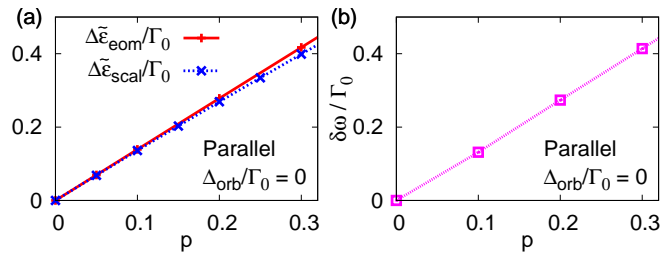}
\end{center}
\caption{\baselineskip=0.77\baselineskip
(Color on-line) (a) Energy splittings between up- and down-spin QD levels as functions of $p$ for the P configuration. $\Delta\tilde{\varepsilon}_{eom}$ and $\Delta\tilde{\varepsilon}_{scal}$ are obtained by the EOM method and the scaling approach, respectively. 
(b) Energy difference $\delta \omega/\Gamma_{0}$ between the peaks of $\rho(\omega)$ in the positively and negatively high-energy region as a function of $p$ for the P configuration. For both (a) and (b), we set $\Delta_{orb}/\Gamma_{0} = 0$, $T/\Gamma_{0} = 0.005$, and $\varepsilon_{0}/\Gamma_{0} = -2$. 
}
\label{fig3}
\end{figure}
We discuss the effects of the ferromagnetic leads in the P configuration on the Kondo effect for $\Delta_{orb}/\Gamma_{0} = 0$. 
In Fig. \ref{fig2}, we show $\rho(\omega) = \sum_{\sigma} \rho_{\sigma}(\omega)$, $\rho_{\uparrow} (\omega)$, and $\rho_{\downarrow} (\omega)$ for several values of the spin polarization $p$. Note that the Fermi energy is $\omega/\Gamma_{0}=0$. Since we find that $\rho_{l \sigma}(\omega)$ are independent of $l$, we set $\rho_{\sigma}(\omega) = \sum_{l} \rho_{l \sigma}(\omega)$. 
In Fig. \ref{fig2}(a), for $p = 0$ one conspicuous peak appears at $\omega/\Gamma_0 \sim 0.02$, 
which results from the spin-orbital symmetric SU(4) Kondo effect \cite{lim06,choi05,sakano06} for $U$ and $U' \gg |\varepsilon_l|$. 
In the SU(4) Kondo system, the Friedel sum rule ensures that the peak of the Kondo resonance appears at $\omega/\Gamma_0 \sim T_K^{SU(4)}/\Gamma_0$, where $T_K^{SU(4)}$ is the Kondo temperature \cite{lim06}. 
From the numerical results, we evaluate $T_K^{SU(4)}/\Gamma_0 \sim 0.02$ for $p=0$ and $\Delta_{orb}/\Gamma_0=0$. 
This value is larger than the used temperature in the calculation $(T/\Gamma_0=0.005)$ and thus the SU(4) Kondo effect appears for $T/\Gamma_0=0.005$.

For $p \neq 0$, the Kondo peak splits into three peaks.  The splittings can be clearly seen for $p \ge 0.1$ as shown in Fig. \ref{fig2}(a). 
The energy difference $\delta \omega/\Gamma_{0}$ between the right and left peaks becomes large with increasing $p$, while the middle peak remains significantly at $\omega/\Gamma_{0} \sim 0$ for finite $p$. 
In Fig. \ref{fig2}(b) and (c), we find that the left and right peaks in $\rho(\omega)$ originate from $\rho_\uparrow(\omega)$ and $\rho_\downarrow(\omega)$, respectively. 
Accordingly, it is considered that for finite $p$ in the P configuration the Kondo effect caused by the spin degrees of freedom (the spin Kondo effect) is suppressed as a result of the local exchange field, while the Kondo effect caused by the orbital degeneracy (the orbital Kondo effect) survives.

We investigate the origin of the splitting in terms of the poor man's scaling approach. 
In Fig. \ref{fig3}(a), we show the spin splitting of the renormalized QD energies ($\Delta\tilde{\varepsilon}_{scal}/\Gamma_{0}$) obtained by the poor man's scaling as a function of $p$. We compare it with the spin splitting of the renormalized QD energies ($\Delta\tilde{\varepsilon}_{eom}/\Gamma_{0}$) obtained by the EOM. According to Eqs. (\ref{eq:se1}) and (\ref{eq:del_e1}), $\Delta\tilde{\varepsilon}_{eom}/\Gamma_{0}$ can be defined as
\begin{eqnarray}
\Delta\tilde{\varepsilon}_{eom} \equiv \Delta\tilde{\varepsilon}^{1}_{l \downarrow} - \Delta\tilde{\varepsilon}^{1}_{l \uparrow}. 
\label{eq:de_eom}
\end{eqnarray}
We find that both calculations are in good quantitative agreement, although $\Delta\tilde{\varepsilon}_{scal}/\Gamma_{0}$ deviates slightly from the linear $p$ dependence of $\Delta\tilde{\varepsilon}_{eom}/\Gamma_{0}$ in $p \gtrsim 0.2$. 
In Fig. \ref{fig3}(b), we show the energy difference $\delta \omega/\Gamma_0$ between the right and left peaks in Fig. \ref{fig2}(a) as a function of $p$. We find that $\delta \omega/\Gamma_0$ is proportional to $p$ and agrees with $\Delta\tilde{\varepsilon}_{eom}/\Gamma_{0}$ quantitatively. 
Therefore, in the P configuration spin-dependent charge fluctuations induce the spin splitting of the renormalized QD energy, which results in the linear $p$ dependence of $\delta \omega/\Gamma_{0}$ shown in Fig. \ref{fig3}(b).

\begin{figure}[t]
\begin{center}
\includegraphics[width=60mm]{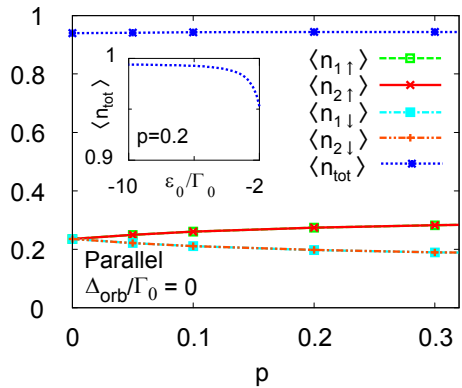}
\end{center}
\caption{\baselineskip=0.77\baselineskip
(Color on-line) QD electron numbers, 
$\langle n_{1\uparrow} \rangle$, $\langle n_{2\uparrow} \rangle$, $\langle n_{1\downarrow} \rangle$, $\langle n_{2\downarrow} \rangle$, and $\langle n_{tot} \rangle$ as functions of $p$ in the P configuration. 
$\langle n_{tot} \rangle = \sum_{l \sigma} \langle n_{l \sigma} \rangle$.
The parameters are $\Delta_{orb}/\Gamma_{0} = 0$, $T/\Gamma_{0} = 0.005$, and $\varepsilon_{0}/\Gamma_{0} = -2$. 
Inset: $\langle n_{tot} \rangle$ as a function of $\varepsilon_{0}/\Gamma_0$ with $p = 0.2$.}
\label{fig4}
\end{figure}
We next investigate the electron number of the QD in the P configuration. 
In Fig. \ref{fig4}, we show $\langle n_{l \sigma} \rangle$ and $\langle n_{tot} \rangle = \sum_{l \sigma} \langle n_{l \sigma} \rangle$ as functions of $p$. 
For $p = 0$, the $\langle n_{l \sigma} \rangle$ values are the same irrespective of $l$ and $\sigma$. 
As $p$ increases, $\langle n_{l \uparrow} \rangle$ increases, while $\langle n_{l \downarrow} \rangle$ decreases. 
For the up-spin (down-spin) electrons, the hybridization $\Gamma_{\uparrow}^1$ ($\Gamma_{\downarrow}^1$) is enhanced (suppressed) with increasing $p$, which results in an increase (decrease) in $\langle n_{l \uparrow} \rangle$ ($\langle n_{l \downarrow} \rangle$). 
On the other hand, we find $\langle n_{1 \sigma} \rangle =\langle n_{2 \sigma} \rangle$ for $p \neq 0$. This result indicates that the orbital degeneracy exists even for finite $p$ in the P configuration, which yields the orbital Kondo effect as shown in Fig. \ref{fig2}(a). Details will be discussed in the next subsection.
The total QD electron number $\langle n_{tot} \rangle$ is almost constant irrespective of $p$ and approaches unity with decreasing $\varepsilon_{0}/\Gamma_0$ as shown in the inset of Fig. \ref{fig4}. This property originates from the strong Coulomb repulsions $U$ and $U' \gg |\varepsilon_l|$.

As shown in the inset of Fig. \ref{fig2}(a), the hump structure around $\omega/\Gamma_{0} =-1$ caused by charge fluctuations is almost independent of $p$. 
The energy at which the hump structure shows its maximum is shifted from the bare QD energy level $\epsilon_0/\Gamma_0=-2$ toward the Fermi energy because of the correlation effects.

\subsection{Orbital splitting effects}
\begin{figure}[t]
\begin{center}
\includegraphics[width=70mm]{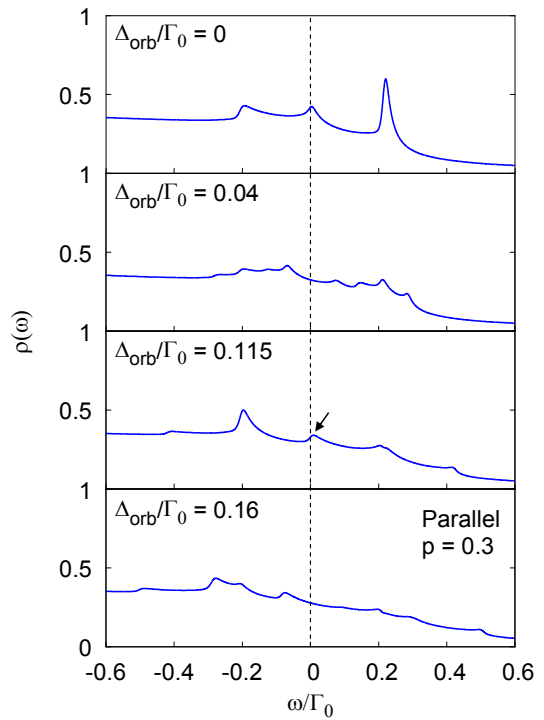}
\end{center}
\caption{\baselineskip=0.77\baselineskip
(Color on-line) Single-particle excitation spectra $\rho (\omega)$ for the several values of $\Delta_{orb}/\Gamma_{0}$ in the P configuration. The parameters are $p = 0.3$, $T/\Gamma_{0} = 0.005$, and $\varepsilon_{0}/\Gamma_{0} = -2$.}
\label{fig5}
\end{figure}
We investigate the effects of the orbital splitting in the P and AP configurations.  
In Fig. \ref{fig5}, we show the SPES in the P configuration for several $\Delta_{orb}/\Gamma_{0}$ values at $p=0.3$. 
For $\Delta_{orb}/\Gamma_{0} = 0$, the Kondo peak caused by orbital fluctuations appears close to $\omega/\Gamma_{0} =0$. 
As $\Delta_{orb}/\Gamma_{0}$ is increased, the peak close to $\omega/\Gamma_{0} =0$ disappears, which indicates that the Kondo effect has vanished. Instead, multipeaks appear. 
At $\Delta_{orb}/\Gamma_{0}=0.115$, we find a small peak at $\omega/\Gamma_{0} \sim 0$ as indicated by the arrow, showing that the Kondo effect is restored. 
When $\Delta_{orb}/\Gamma_{0}$ is further increased, the peak close to $\omega/\Gamma_{0} =0$ disappears again and the multipeaks appear.

\begin{figure}[t]
\begin{center}
\includegraphics[width=60mm, angle=0]{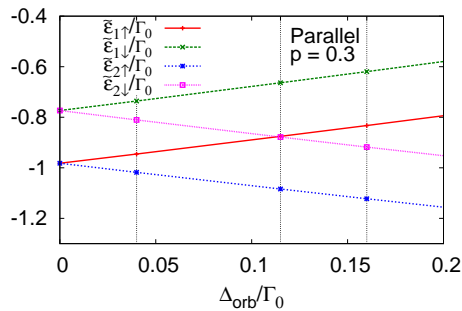}
\end{center}
\caption{\baselineskip=0.77\baselineskip
(Color on-line) Renormalized QD energy levels in the P configuration as functions of $\Delta_{orb}/\Gamma_{0}$ for $p = 0.3$. The vertical dashed lines are $\Delta_{orb}/\Gamma_{0}=0.04$, $0.115$, and $0.16$, which correspond to the values used in Fig. \ref{fig5}. $\tilde{\varepsilon}_{l \uparrow}/\Gamma_{0}$ and $\tilde{\varepsilon}_{2 \downarrow}/\Gamma_{0}$ cross at $\Delta_{orb}/\Gamma_{0}=0.115$.}
\label{fig6}
\end{figure}
The peak positions in $\rho(\omega)$ are determined by the $\omega$ dependence of the self-energies $\Sigma^1_{l \sigma} (\omega)$, $\Sigma^2_{l \sigma} (\omega)$, and $\Sigma^3_{l \sigma} (\omega)$, which reflect $\Delta \tilde{\varepsilon}_{l \sigma}^{1}$, $\Delta \tilde{\varepsilon}_{l \sigma}^{2}$, and $\Delta \tilde{\varepsilon}_{l \sigma}^{3}$. 
We show $\tilde{\varepsilon}_{l \sigma}/\Gamma_{0}$ as functions of $\Delta_{orb}/\Gamma_{0}$ in Fig. \ref{fig6}. 
The vertical dashed lines are $\Delta_{orb}/\Gamma_{0}=0.04$, $0.115$, and $0.16$, which we have used in Fig. \ref{fig5}.

For $\Delta_{orb}/\Gamma_{0}=0$, we find $\tilde{\varepsilon}_{1 \sigma}/\Gamma_{0}=\tilde{\varepsilon}_{2 \sigma}/\Gamma_{0}$ and $\tilde{\varepsilon}_{l \uparrow}/\Gamma_{0} \neq \tilde{\varepsilon}_{l \downarrow}/\Gamma_{0}$. According to Eqs. (\ref{eq:del_e1}), (\ref{eq:del_e2}), and (\ref{eq:del_e3}), $\Delta \tilde{\varepsilon}_{l \sigma}^{1}$ and $\Delta \tilde{\varepsilon}_{l \sigma}^{3}$ have two different values, and $\Delta \tilde{\varepsilon}_{l \sigma}^{2}=0$. Because of these properties, for $\Delta_{orb}/\Gamma_{0}=0$, the Kondo peak caused by orbital fluctuations appears and the splitting two peaks appear in $\rho(\omega)$. 
As $\Delta_{orb}/\Gamma_{0}$ is increased, the orbital degeneracy is lifted and $\tilde{\varepsilon}_{1 \sigma}/\Gamma_{0}$ ($\tilde{\varepsilon}_{2 \sigma}/\Gamma_{0}$) increase (decrease) with the same $\Delta_{orb}/\Gamma_{0}$ dependence as shown in Fig. \ref{fig6}. 
Accordingly, for $\Delta_{orb}/\Gamma_{0} \neq 0$, $\Delta \tilde{\varepsilon}_{l \sigma}^{1}$ and $\Delta \tilde{\varepsilon}_{l \sigma}^{2}$ exhibit two different values depending on $\sigma$ and $l$, respectively, and $\Delta \tilde{\varepsilon}_{l \sigma}^{3}$ exhibits four different values. 
Because of these features, the Kondo effect disappears for $\Delta_{orb}/\Gamma_{0}=0.04$ and $0.16$, and eight peaks appear in $\rho(\omega)$ as shown in Fig. \ref{fig5}.

For $\Delta_{orb}/\Gamma_{0}=0.115$, we find a noticeable feature: $\tilde{\varepsilon}_{1 \uparrow}/\Gamma_{0}$ and $\tilde{\varepsilon}_{2 \downarrow}/\Gamma_{0}$ cross as shown in Fig. \ref{fig6} and new two-fold degenerate states arise there. 
The scattering between these two degenerate states can be expressed using the pseudospin-flip process. This process is introduced via the self-energy $\Sigma^3_{l \sigma} (\omega)$, resulting in an unconventional Kondo effect. 
The small peak close to $\omega/\Gamma_{0} =0$ indicated by the arrow in Fig. \ref{fig5} is a Kondo resonance caused by this scattering process. 
Since the respective gradients of $\tilde{\epsilon}_{1 \sigma}/\Gamma_{0}$ and $\tilde{\epsilon}_{2 \sigma}/\Gamma_{0}$ are the same, we find only one crossing point as a function of $\Delta_{orb}/\Gamma_0$. Although the $\Delta_{orb}/\Gamma_0$ value at the crossing point depends on the parameters, the result does not change qualitatively: The new degeneracy arises at a certain value of $\Delta_{orb}/\Gamma_0$, which can be evaluated from the magnitude of the energy splitting $\tilde{\varepsilon}_{l\downarrow}-\tilde{\epsilon}_{l\uparrow}$ at $\Delta_{orb}=0$, and the gradients of $\tilde{\varepsilon}_{l\sigma}/\Gamma_{0}$ as a function of $\Delta_{orb}$.
The similar situation has been observed as a singlet-triplet Kondo effect and a doublet-doublet Kondo effect \cite{sasaki2005,eto2005}.
It is considered that this Kondo effect can be observed in the field dependence of the conductance, which will be described in the next subsection.

\begin{figure}[t]
\begin{center}
\includegraphics[width=70mm]{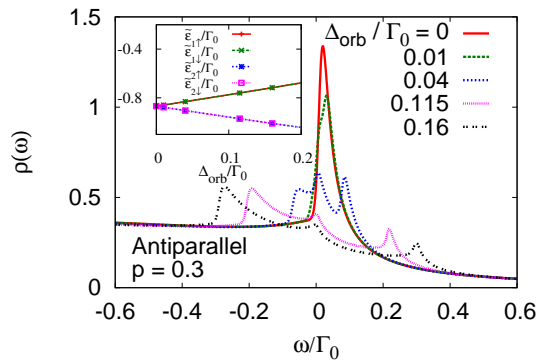}
\end{center}
\caption{\baselineskip=0.77\baselineskip
(Color on-line) Single-particle excitation spectra $\rho (\omega)$ for the several values of $\Delta_{orb}/\Gamma_{0}$ in the AP configuration. The parameters are $\Delta_{orb}/\Gamma_{0} = 0$, $p = 0.3$, $T/\Gamma_{0} = 0.005$, and $\varepsilon_{0}/\Gamma_{0} = -2$. 
Inset: Renormalized QD energy levels in the AP configuration as functions of $\Delta_{orb}/\Gamma_{0}$ for $p = 0.3$. }
\label{fig7} 
\end{figure}
The results for the AP configuration are shown in Fig. \ref{fig7}. For $\Delta_{orb}/\Gamma_{0}=0$, the spectral profile is the same as that of the SU(4) Kondo effect shown in Fig. \ref{fig2}(a). Therefore, the effects of the ferromagnetic leads vanish in the AP configuration for a zero magnetic field. 
For $\Delta_{orb}/\Gamma_{0}=0.01$, we find a single peak and the SU(4) Kondo effect remains there. 
As $\Delta_{orb}/\Gamma_{0}$ is further increased, the SU(4) Kondo peak splits into three peaks. The peaks at either end move further apart with increasing $\Delta_{orb}/\Gamma_{0}$, while the middle peak remains at $\omega/\Gamma_0 \sim 0$ irrespective of $\Delta_{orb}/\Gamma_{0}$. 
In the inset of Fig. \ref{fig7}, we show the renormalized QD energy levels as a function of $\Delta_{orb}/\Gamma_{0}$ for $p = 0.3$. We find that $\tilde{\varepsilon}_{1 \uparrow}/\Gamma_{0}=\tilde{\varepsilon}_{1 \downarrow}/\Gamma_{0}$ and $\tilde{\varepsilon}_{2 \uparrow}/\Gamma_{0}=\tilde{\varepsilon}_{2 \downarrow}/\Gamma_{0}$, and that the two levels $\tilde{\varepsilon}_{1 \sigma}/\Gamma_{0}$ and $\tilde{\varepsilon}_{2 \sigma}/\Gamma_{0}$ deviate with increasing $\Delta_{orb}/\Gamma_{0}$. 
The results indicate that in the AP configuration the orbital Kondo effect is suppressed for finite $\Delta_{orb}/\Gamma_{0}$, while the spin Kondo effect remains.

\subsection{Conductance}
\begin{figure}[t]
\begin{center}
\includegraphics[width=70mm]{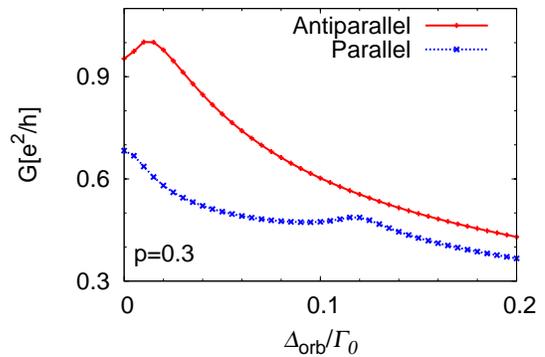}
\end{center}
\caption{\baselineskip=0.77\baselineskip
(Color on-line) Conductance as functions of $\Delta_{orb}/\Gamma_{0}$ for $p = 0.3$ in the P and AP configurations. 
We set $T/\Gamma_{0}=0.005$. 
The local maximum appears at $\Delta_{orb}/\Gamma_{0} \sim 0.115$ in the P configuration. 
}
\label{fig8}
\end{figure}
Substituting $\rho(\omega)$ for various $\Delta_{orb}/\Gamma_{0}$ values into Eq. (\ref{eq:G}), we obtain the $\Delta_{orb}/\Gamma_{0}$ dependence of the conductance $G$. The results for $p=0.3$ in the P and AP configurations are shown in Fig. \ref{fig8}. 
In the P configuration, the conductance first decreases with increasing $\Delta_{orb}/\Gamma_{0}$ because of the suppression of the Kondo effect. As $\Delta_{orb}/\Gamma_{0}$ increases further, the conductance increases nontrivially and takes a local maximum at $\Delta_{orb}/\Gamma_{0} \sim 0.115$. This nontrivial local maximum is caused by the Kondo effect indicated by the arrow in Fig. \ref{fig5}. 
By contrast, in the AP configuration, the conductance has its maximum at $\Delta_{orb}/\Gamma_0 \sim 0.01$. We discuss the origin of this maximum. 
For $\Delta_{orb}/\Gamma_0 \lesssim 0.01$, the SU(4) Kondo effect remains as shown in Fig. \ref{fig7}. 
However, the spectral weight close to the Fermi energy at $\Delta_{orb}/\Gamma_0 =0.01$ becomes slightly larger than that at $\Delta_{orb}/\Gamma_0 =0$ as shown in Fig. \ref{fig7}. 
For further increase in $\Delta_{orb}/\Gamma_0$, the orbital Kondo effect is suppressed. The contributions of two spectral peaks at both ends to the spectral weight at $\omega/\Gamma_0 \sim 0$ are reduced with increasing $\Delta_{orb}/\Gamma_0$, leading to a monotonic decrease in $G$. Accordingly, the conductance has a maximum at $\Delta_{orb}/\Gamma_0 \sim 0.01$.

We now discuss the effects of the polarization $p$ on the Kondo temperature. 
In the AP configuration, the SU(4) Kondo effects at $\Delta_{orb}/\Gamma_0=0$ is considered to appear even for finite $p$ and $T_K^{\rm SU(4)}$ is independent of $p$. On the other hand, when $p$ is introduced for $\Delta_{orb}=0$ in the P configuration, the SU(4) Kondo effect is destroyed and the SU(2) orbital Kondo effect survives as shown in Fig. \ref{fig2}(a). We evaluate the Kondo temperature from the inflection point of the temperature dependence of G in the P configuration (not shown in Figure). 
For $p=0$, we obtain $T_K^{\rm SU(4)}/\Gamma_0 \sim 0.02$ for the SU(4) Kondo effect. This value is consistent with that obtained from the peak energy of the Kondo resonance shown in Fig. \ref{fig2}(a). For $p=0.3$, we obtain $T_K^{\rm SU(2)}/\Gamma_0 \sim 0.005$ for the SU(2) Kondo effect. 
These results are reasonable, because $T_K^{\rm SU(4)}$ is expected to be higher than $T_K^{\rm SU(2)}$. 
Accordingly, the effective temperature scaled by $T_K^{\rm SU(4)}$ for $p=0$ is lower than that scaled by $T_K^{\rm SU(2)}$ for $p=0.3$ in the P configuration. This result qualitatively explains the conductance at $\Delta_{orb}/\Gamma_0=0$ shown in Fig. \ref{fig8}, because in the AP configuration the SU(4) Kondo effect appears at $\Delta_{orb}=0$: 
At $\Delta_{orb}/\Gamma_0=0$, the effective temperature in the P configuration is higher than that in the AP configuration, which yields the smaller G in the P configuration.
It is known that the EOM tends to underestimate the spectral weights and thus the magnitude of the conductance \cite{ferroEOM1,polish06}. Actually, the conductance for $p=0$ and $\Delta_{orb}=0$ does not approach the value in the unitary limit $2e^2/h$ for the SU(4) Kondo effect. To obtain more quantitatively accurate results, another numerical calculation by, for example, a numerical renormalization-group may be effective. Such a research is our future study.

Finally, we discuss the conductance of a vertical QD coupled to two Ni ferromagnetic leads. To develop a quantitative discussion, we use $\Gamma_0=30$ ${\rm meV}$, which was estimated in a ${\rm C_{60}}$ experiment \cite{ferroC60}. The range of $\Delta_{orb}/\Gamma_{0}=0.2$ in Fig. \ref{fig8} thus corresponds to $\Delta_{orb}=6$ ${\rm meV}$. According to the Fock-Darwin state with an electron confinement energy of $3$ ${\rm meV}$ \cite{tarucha96-97}, we evaluate that $\Delta_{orb}=6$ ${\rm meV}$ corresponds to the magnetic field $B \sim 3.5 {\rm T}$. 
On the basis of these results, we discuss the change in the conductance in external magnetic fields \cite{comm}. 
We first set the QD coupled to the Ni ferromagnetic leads in the AP configuration. The conductance in a zero magnetic field is $G|_{B=0} \sim 0.952$ ${\rm e^2/h}$ as shown in Fig. \ref{fig8}. When the magnetic field is increased, the AP configuration changes into the P configuration at $B \sim O(10^{-1}) {\rm T}$ \cite{ferroC60}. When the magnetic field is further increased, we observe a nontrivial local maximum at $B \sim 1.9 {\rm T}$ and then obtain $G|_{B=3.5} \sim 0.367$ ${\rm e^2/h}$ at $B \sim 3.5 {\rm T}$. 
We evaluate the ratio of the change in $G$ as $\Delta G \equiv \left( G|_{B=0}-G|_{B=3.5{\rm T}} \right)/G|_{B=3.5{\rm T}} \sim 1.59$. This value is nearly twice that in the single-orbital QD coupled to the Ni ferromagnetic leads, where the change ratio in $G$ between the AP and P configurations is estimated to be $\Delta G \sim 0.81$. Note that $\Delta G$ has the same meaning as tunnel magnetoresistance. 
We have demonstrated that the large gain in $\Delta G$ can be controlled by an external magnetic field in a two-orbital vertical QD coupled to two ferromagnetic leads.

\section{Summary} \label{sec:sum}
We have investigated the Kondo effect of a two-orbital vertical QD coupled to two ferromagnetic leads using the EOM method. We have shown that in the P configuration the orbital Kondo effect remains, while the spin Kondo effect is suppressed. In magnetic fields, the orbital Kondo effect also disappears, because the energy levels of the QD are completely split. However, at a certain magnetic field, two of the four energy levels cross and the Kondo effect newly emerges there. This Kondo effect can be possibly observed in experiments. 
In the AP configuration the spin Kondo effect remains, while the orbital Kondo effect is suppressed in magnetic fields. 
We have demonstrated that the change ratio in the conductance between the AP and P configurations can be controlled by an external magnetic field. 
In a typical two-orbital QD coupled to Ni ferromagnetic leads, the change ratio of the conductance is larger than that in a single-orbital QD system.

\section*{Acknowledgments}
Some of the numerical computations were performed at the Supercomputer Center at 
ISSP, University of Tokyo. T.K. was supported by the Japan Society for 
the Promotion of Science.
This work was supported by Grants-in-Aid for Scientific Research (C)
(No. 20540390) from the Japan Society for the Promotion of Science and on Innovative Areas (No. 21104514) from the Ministry of Education, Culture, Sports, Science and Technology.

\appendix
\section{Derivation of Eqs. (\ref{eq:gf})-(\ref{eq:se3})}
We begin our discussion with the retarded Green's function $G_{l \sigma}^r (t) = - i \theta (t) \langle \{ d_{l \sigma} (t), d_{l \sigma}^\dagger (0) \} \rangle$. In the first iteration of the EOM procedure, we obtain
\begin{eqnarray}
i\frac{d}{dt}G_{l \sigma}^{r}(t) &=& \delta(t) + \varepsilon_lG_{l \sigma}^r (t) + \sum_{i=L,R} \sum_{k} V^{*}_{ikl\sigma} G_{ik,l\sigma}^r (t) \nonumber\\
&+& U\Pi_{l \sigma}(t) + U'\Lambda_{l \sigma}(t), 
\label{eq:it1}
\end{eqnarray}
where $G_{ik,l\sigma}^r (t) \equiv - i\theta (t) \langle \{ c_{ik l\sigma}(t), d_{l \sigma}^{\dagger}(0) \} \rangle$, 
$\Pi_{l \sigma}(t) \equiv - i \theta (t) \langle \{ n_{l \overline{\sigma}}(t) d_{l \sigma}(t), d_{l \sigma}^{\dagger}(0) \} \rangle$, and 
$\Lambda_{l \sigma}(t) \equiv - i \theta(t) \sum_{m(\neq l) \alpha} \langle \{ n_{m \alpha}(t) d_{l \sigma}(t), d_{l \sigma}^{\dagger}(0) \} \rangle$. 
We perform the second iterations of the EOM procedure for $G_{ik,l\sigma}^r (t)$, $\Pi_{l \sigma}(t)$, and $\Lambda_{l \sigma}(t)$. 
For $G_{ik,l\sigma}^r (t)$, we obtain the closed form as $idG_{ik,l\sigma}^r (t)/dt = \varepsilon_{ik\sigma}G_{ik,l\sigma}^r (t) + V_{ikl\sigma} G_{l \sigma}^r (t)$. For the latter two ones, the higher-order Green's functions appear as 
\begin{widetext}
\begin{eqnarray}
i\frac{d}{dt}\Pi_{l \sigma}(t)&=& 
\langle n_{l\overline{\sigma}} \rangle \delta(t) 
+ \varepsilon_l\Pi_{l \sigma}(t) 
- \sum_{i=L,R} \sum_{k} V_{ikl\sigma} X_{ik,l\sigma}^A(t)
- \sum_{i=L,R} \sum_{k} V^{*}_{ikl\sigma} X_{ik,l\sigma}^B(t)
+ \sum_{i=L,R} \sum_{k} V^{*}_{ikl\sigma} X_{ik,l\sigma}^C(t)   \nonumber\\
&+& U\Pi_{l \sigma}(t) 
+ U'\sum_{m(\neq l)\alpha} W_{m\alpha,l\sigma}(t), 
\label{eq:it2-1}
\end{eqnarray}
\begin{eqnarray}
i\frac{d}{dt}\Lambda_{l \sigma}(t) &=& 
\sum_{m(\neq l)\alpha} \langle n_{m\alpha} \rangle \delta(t) 
+ \varepsilon_l\Lambda_{l \sigma}(t) 
- \sum_{i=L,R}\sum_{k}\sum_{m(\neq l)\alpha} V_{ikl\sigma} Y_{ikm\alpha,l\sigma}^A(t)  - \sum_{i=L,R}\sum_{k}\sum_{m(\neq l)\alpha} V^{*}_{ikl\sigma} Y_{ikm\alpha,l\sigma}^B(t) \nonumber\\
&+& \sum_{i=L,R}\sum_{k}\sum_{m(\neq l)\alpha} V^{*}_{ikl\sigma} Y_{ikm\alpha,l\sigma}^C(t)
+ U\sum_{m(\neq l)\alpha} W_{m\alpha,l\sigma}(t)
+U'\sum_{m(\neq l)\alpha}\sum_{n(\neq l)\beta}Z_{m\alpha n\beta,l\sigma}(t),
\label{eq:it2-2}
\end{eqnarray}
\end{widetext}
where $\langle n_{l \sigma} \rangle$ is the electron number of the QD with spin $\sigma$ and orbital $l$, and 
$X_{ik,l\sigma}^A(t) \equiv -i\theta(t) \langle \{ c_{ikl\overline{\sigma}}^\dagger(t) d_{l\overline{\sigma}}(t) d_{l \sigma}(t), d_{l \sigma}^\dagger(0) \} \rangle$, 
$X_{ik,l\sigma}^B(t) \equiv -i\theta(t) \langle \{ d_{l\overline{\sigma}}^\dagger(t) d_{l \sigma}(t) c_{ikl\overline{\sigma}}(t), d_{l \sigma}^\dagger(0) \} \rangle$, 
$X_{ik,l\sigma}^C(t) \equiv -i\theta(t) \langle \{ n_{l\overline{\sigma}}(t) c_{ikl\sigma}(t), d_{l \sigma}^\dagger(0) \} \rangle$, 
$W_{m\alpha,l\sigma}(t) \equiv -i\theta(t) \langle \{ n_{m\alpha}(t) n_{l\overline{\sigma}}(t) d_{l \sigma}(t), d_{l \sigma}^\dagger(0) \} \rangle$, 
$Y_{ikm\alpha,l\sigma}^A(t) \equiv -i\theta(t) \langle \{ c_{ikm\alpha}^\dagger(t) d_{m\alpha}(t) d_{l \sigma}(t), d_{l \sigma}^\dagger(0) \} \rangle$, 
$Y_{ikm\alpha,l\sigma}^B(t) \equiv -i\theta(t) \langle \{ d_{m\alpha}^\dagger(t) d_{l \sigma}(t) c_{ikm\alpha}(t), d_{l \sigma}^\dagger(0) \} \rangle$, 
$Y_{ikm\alpha,l\sigma}^C(t) \equiv -i\theta(t) \langle \{ n_{m\alpha}(t) c_{ikl\sigma}(t), d_{l \sigma}^\dagger(0) \} \rangle$, 
$Z_{m\alpha n\beta,l\sigma}(t) \equiv -i\theta(t) \langle \{ n_{m\alpha}(t) n_{n\beta}(t) d_{l \sigma}(t), d_{l \sigma}^\dagger(0) \} \rangle$.

We perform the third iterations of the EOM procedure for these higher-order Green's functions which appear in the right-hand side of Eqs. (\ref{eq:it2-1}) and (\ref{eq:it2-2}). In this calculation, we adopt the approximation which successfully yields the closed equation of the Green's functions in a single-orbital QD system \cite{EOM1,EOM2}. According to this approximation, we neglect the following types of higher-order terms 
$ \langle \{ c_{i k l \overline{\sigma}}^\dagger(t) d_{l \overline{\sigma}}(t) c_{i^\prime k^\prime l \sigma}(t), d_{l \sigma}^\dagger(0) \} \rangle$ and 
$\langle \{ d_{l \overline{\sigma}}^\dagger(t) c_{i k l \overline{\sigma}} (t) c_{i^\prime k^\prime l \sigma} (t) , d_{l \sigma} (0)^\dagger \} \rangle$ as well as the expectation values 
$\langle c_{i k l \sigma}^\dagger d_{l \sigma} \rangle$ and $\langle d_{l \sigma}^\dagger c_{i k l \sigma} \rangle$, and decouple 
$\langle \{ c_{i k l \overline{\sigma}} (t)^\dagger c_{i^\prime k^\prime l \overline{\sigma}} (t) d_{l \sigma} (t), d_{l \sigma} (0)^\dagger \} \rangle \sim \delta_{i i^\prime} \delta_{k k^\prime} f(\varepsilon_{ik\overline{\sigma}}) \langle \{d_{l \sigma} (t) , d_{l \sigma} (0)^\dagger \} \rangle$. 
In addition to this approximation, we decouple the number operator of the localized electron and replace it by its expectation value. 
Furthermore, we turn our attention to the scattering processes of the electrons in the same orbital with different spins, and in the different orbitals with the same and different spins. We include the higher-order terms which express these scattering processes and neglect those expressing other higher-order scattering processes. 
Using these approximations, we obtain the following equations of the higher-order Green's functions 
\begin{widetext}
\begin{eqnarray}
i\frac{d}{dt}X_{ik,l\sigma}^A(t)= 
-\left( \varepsilon_{ik\overline{\sigma}}-2\varepsilon_{l}-U-
2U'\sum_{m(\neq l)\alpha}\langle n_{m\alpha}\rangle \right) 
X_{ik,l\sigma}^A(t) 
-V^{*}_{ikl\sigma}\Pi_{l \sigma}(t) + V^{*}_{ikl\sigma}
f(\varepsilon_{ik\overline{\sigma}})G_{l \sigma}^{r}(t),
\label{eq:it3-1}
\end{eqnarray}
\begin{eqnarray}
i\frac{d}{dt}X_{ik,l\sigma}^B(t)= 
\varepsilon_{ik\overline{\sigma}}X_{ik,l\sigma}^B(t) +  V^{*}_{ikl\sigma}
f(\varepsilon_{ik\overline{\sigma}})G_{l \sigma}^{r}(t)
-V_{ikl\sigma}\Pi_{l \sigma}(t),
\label{eq:it3-2}
\end{eqnarray}
\begin{eqnarray}
i\frac{d}{dt}X_{ik,l\sigma}^C(t)= 
\varepsilon_{ik\sigma}X_{ik,l\sigma}^C(t) + V_{ikl\sigma}\Pi_{l \sigma}(t),
\label{eq:it3-3}
\end{eqnarray}
\begin{eqnarray}
i\frac{d}{dt}W_{m\alpha,l\sigma}(t) &=& 
\langle n_{m\alpha} \rangle \langle n_{l\overline{\sigma}} \rangle \delta(t) 
+\left( \varepsilon_{l}+U+U'\sum_{j(\neq l)s}\langle n_{js}\rangle \right)
W_{m\alpha,l\sigma}(t) 
- \sum_{k}V_{ikl\sigma}\langle n_{m\alpha}\rangle X_{ik,l\sigma}^A(t) 
\nonumber\\ 
&-& \sum_{k}V^{*}_{ikl\sigma}\langle n_{m\alpha}\rangle X_{ik,l\sigma}^B(t),
\label{eq:it3-4}
\end{eqnarray}
\begin{eqnarray}
i\frac{d}{dt}Y_{ik,l\sigma}^A(t)= 
-\left( \varepsilon_{ik\overline{\sigma}}-\varepsilon_{l}-\varepsilon_{m}+U'
+U'\sum_{j(\neq l)s} \langle n_{js}\rangle + 
U\langle n_{l\overline{\sigma}}\rangle \right) Y_{ik,l\sigma}^A(t) 
+\left[ f(\varepsilon_{ik\overline{\sigma}})-\langle n_{m\alpha}\rangle \right]
V^{*}_{ikl\sigma}G_{l \sigma}^{r}(t),
\label{eq:it3-5}
\end{eqnarray}
\begin{eqnarray}
i\frac{d}{dt}Y_{ik,l\sigma}^B(t)= 
\left( \varepsilon_{ik\overline{\sigma}}+\varepsilon_{l}-\varepsilon_{m}-U'
+U'\sum_{j(\neq l)s} \langle n_{js}\rangle + 
U\langle n_{l\overline{\sigma}}\rangle \right) Y_{ik,l\sigma}^B(t) 
+\left[ f(\varepsilon_{ik\overline{\sigma}})-\langle n_{m\alpha}\rangle \right]
V_{ikl\sigma}G_{l \sigma}^{r}(t),
\label{eq:it3-6}
\end{eqnarray}
\begin{eqnarray}
i\frac{d}{dt}Y_{ik,l\sigma}^C(t)= 
\varepsilon_{ik\sigma}Y_{ik,l\sigma}^C(t) 
+\langle n_{m\alpha}\rangle V_{ikl\sigma}G_{l \sigma}^{r}(t),
\label{eq:it3-7}
\end{eqnarray}
\begin{eqnarray}
i\frac{d}{dt}Z_{m\alpha n\beta,l\sigma}(t) = 
\langle n_{m\alpha} \rangle \delta_{m,n} \delta_{\alpha,\beta} \delta(t)
+\left( \varepsilon_{l} + U\langle n_{l\overline{\sigma}}\rangle 
+U'\langle n_{l\overline{\sigma}}\rangle \right)Z_{m\alpha n\beta,l\sigma}(t). 
\label{eq:it3-8}
\end{eqnarray}
\end{widetext}
We solve these coupled differential equations by the Fourier transformation with respect to $t$. 
In the condition of the strong Coulomb interactions $U$ and $U^{\prime} \rightarrow \infty$ which corresponds to $U$ and $U' \gg |\varepsilon_l|$, we derive the equations for the Green's function as shown in Eqs. (\ref{eq:gf})-(\ref{eq:se3}).


\begin{thebibliography}{99}
\bibitem{comm}
Using these parameters, we estimate the Zeeman energy as $E_Z/\Gamma_{0} \sim O(10^{-4})$. This value is smaller than the Kondo temperature $T_K/\Gamma_{0} \sim O(10^{-2})-O(10^{-3})$ and the orbital splitting $\Delta_{orb}/\Gamma_{0} \sim O(10^{-1})$. Therefore, we can ignore the Zeeman effect.
\end{thebibliography}
\end{document}